\documentclass[showpacs,preprintnumbers,amsmath,amssymb,nofootinbib]{revtex4}

\usepackage{graphicx}
\usepackage{dcolumn}
\usepackage{bm}

\usepackage[utf8]{inputenc}
\usepackage[T1]{fontenc}

\begin{document}

\def\bb    #1{\hbox{\boldmath${#1}$}}

\title{ The Compactification of QCD$_4$ to QCD$_2$ in a Flux Tube}

\author{Andrew V. Koshelkin}

\affiliation{Moscow Institute for Physics and Engineering,
Kashirskoye shosse, 31, 115409 Moscow, Russia}

\author{Cheuk-Yin Wong}

\affiliation{Physics Division, Oak Ridge National Laboratory, Oak
Ridge, TN
%\footnote{wongc@ornl.gov}
37831}

\date{\today}

\begin{abstract}

We show from the action integral that in the special environment
of a flux tube, QCD$_4$ in (3+1) dimensional space-time can be
approximately compactified into QCD$_2$ in (1+1) dimensional
space-time.  In such a process, we find out how the coupling
constant $g_{2D}$ in QCD$_2$ is related to the coupling constant
$g_{4D}$ in QCD$_4$.  We show how the quark and the gluon in QCD$_2$
acquire contributions to their masses arising from their
confinement within the tube, and how all these quantities depend
on the excitation of the partons in the transverse degrees of
freedom.  The compactification facilitates the investigation of
some dynamical problems in QCD$_4$ in the simpler dynamics of QCD$_2$
where the variation of the gluon fields leads to a bound state.

\end{abstract}


\maketitle

\section{Introduction}

Previously, t'Hooft showed that in the limit of large $N_c$ with
fixed $g^2N_c$ in single-flavor QCD$_4$, planar diagrams with
quarks at the edges dominate, whereas diagrams with the topology
of a fermion loop or a wormhole are associated with suppressing
factors of $1/N_c$ and $1/N_c^2$, respectively \cite{tho74a}.  In
this case a simple-minded perturbation expansion with respect to
the coupling constant $g$ cannot describe the spectrum, while the
$1/N_c$ expansion may be a reasonable concept, in spite of the
fact that $N_c$ is equal to 3 and is not very big.  The dominance
of the planar diagram allows one to consider QCD in one space and
one time dimensions (QCD$_2$) and the physics resembles those of the
dual string or a flux tube, with the physical spectrum of a
straight Regge trajectory \cite{tho74b}.  Since the pioneering
work of t'Hooft, the properties of QCD in two-dimensional
space-time have been investigated by many workers
\cite{tho74a,tho74b,Fri93,Dal93,Fri94,Arm95,Kut95,Gro96,Abd96,Dal98,Arm99,Eng01,Tri02,Abr04,Li87,Wit84}.

The flux tube picture of longitudinal dynamics is
phenomenologically supported in hadron spectroscopy \cite{Isg85},
in hadron collisions, and in $e^+e^-$ annihilations at high
energies \cite{Cas74,And83,Won91,Gat92,Won94,Won09,Won10}.  In
these high-energy processes, the average transverse momenta of
produced hadrons are observed to be limited, of the order of a few
hundred MeV. In contrast, the longitudinal momenta of the produced
hadrons can be very large, as described by a rapidity plateau with
a large average longitudinal momentum.   This average longitudinal
momentum increases with the collision energy.  The limitation of
the average transverse momenta of the produced hadrons means that
the average momenta of partons in produced hadrons are also
limited,\footnote{Even though the
  average transverse momenta of the partons are limited, the tails of
  the parton transverse momentum distribution of partons in the
  produced hadrons can still extend to the high $p_T$ region, but with
  small probabilities.}  consistent with the picture that the produced
partons as constituents of the produced hadrons are transversely
confined in a flux tube.  Further idealization of the
three-dimensional flux tube as a one-dimensional string leads to
the picture of the particle production process as a string
fragmentation in (1+1) space-time dimensions.  The particle
production description of Casher, Kogut, and Susskind \cite{Cas74}
in (1+1) dimensional Abelian gauge theory led to results that
mimics the dynamics of particle production in hadron collisions
and in the annihilation of $e^+e^-$ pairs at high energies.
Furthermore, the Lund model of classical string fragmentation has
been quite successful in describing quantitatively the process of
particle production in these high energy processes
\cite{And83,Won94}.

With the successes of the theoretical description of Casher
$et~al.$ and the Lund model of string fragmentation, it should be
possible to compactify quantum chromodynamics in (3+1) dimensional
space-time (QCD$_4$) approximately to quantum chromodynamics in
(1+1) dimensional space-time (QCD$_2$), in the special environment
appropriate for particle production at high-energies.  It is
useful to examine the circumstances under which such a
compactification is possible. Such a link was given earlier in
\cite{Won09, Won10} and reported briefly in \cite{And12}.  Here,
we would like to examine the problem from the more general
viewpoint of the action integral.

We note that the process of string fragmentation  occurs when a
valence quark-antiquark pair pull part from each other at high
energies, as described in \cite{Cas74,And83}.  It is therefore
reasonable to examine the QCD$_4$ compactification under the
dominance of longitudinal dynamics in the center-of-mass frame of
the receding valence $q \bar q$ pair. Under such a longitudinal
dominance in this frame, not only are the magnitudes of the
longitudinal momenta of the leading valence quark and antiquark
dominant over their transverse momenta, so too are the magnitudes
of longitudinal momenta of the produced $q\bar q$ parton pairs.
The spatially one-dimensional string is an idealization of a more
realistic three-dimensional flux-tube.  The description of
produced $q\bar q$ parton pairs residing within the string or flux
tube presumes the confinement of these produced partons in the
string. Hence, it is reasonable to examine further the QCD$_4$
compactification under transverse confinement. As transverse
confinement is a nonperturbative process and is beyond the realm
of perturbative QCD, we can describe the transverse confinement
property in terms of a confining scalar interaction $S({\bb
r}_\perp)$ in transverse coordinates $\bb r_\perp$, with the quark
mass function described by $m(\bb r_\perp)$=$m_0$+$S({\bb
r}_\perp)$ where $m_0$ is the quark rest mass.

Having spelled out explicitly the circumstances under which the
QCD$_4$ compactification may occur, we proceed to start with the
QCD$_4$ action integral and begin our process of compactification.
We need to find out how we can relate the field variables in
four-dimensional space-time to those in 2-dimensional space-time
in such a way that the four-dimensional action integral can be
simplified to contain only field quantities in two-dimensional
space-time.  What is the form of the two-dimensional action
integral after compactification? How are the coupling constant
$g_{2D}$ in the two-dimensional action integral related to the
coupling constant $g$=$g_{4D}$ in QCD$_4$ in four-dimensional
space-time?  Are there additional terms in the two-dimensional
action integral that arise from the compactification? How do all
these quantities depend on the excitation of the partons in the
transverse degrees of freedom?

We shall show that the compactification for QCD$_4$ in a flux tube
leads to an action integral of a QCD gauge field coupled to the
quark field in two-dimensional space time, which can be
appropriately called QCD$_2$. The QCD gauge field coupling constant
is found to depend on the quark transverse wave function in the
flux tube.  There are additional quark- and gluon-mass terms that
arise from the confinement of the quark and the gluon within the
tube.

The success of the compactification program facilitates the
examination of some problems in QCD$_4$ in the simpler dynamics of
QCD$_2$. The QCD$_2$ action integral allows one to obtain the equations
of motion for the quark field and the gauge field.  We find
self-consistent solution of a boson state with a mass in the flux
tube environment, similar to Schwinger's solution of a massive
boson in two-dimensional Abelian gauge field theory.

It should be noted that the occurrence of a massive composite
bound state in gauge field theories has been known in many
previous investigations \cite{xxx}.  While the basic principles of
the massive bound state as arising from interactions of the gauge
fields  in these theories are the same as in the present
investigation in a flux tube, the physical environments and the
constraints are quite different. How the massive boson in a flux
tube environment examined here can be related to the massive boson
formed by purely gluons as a pole in the three gluon vertex in
4-dimensional space-time \cite{Pap11,Pap12} is a subject worthy of
further investigation.

This paper is organized as follows. In Sec.\ II, we show how the
action integral in  QCD$_4$ can be compactified into  QCD$_2$, under
the assumption of longitudinal dominance and transverse
confinement. The relationship between the 4-dimensional (4D)
quantities and those two-dimensional counterpart are expressed
explicitly. The fermions and gauge bosons acquire contributions to
their masses that arises from the confinement.  In Sec.\ III, we
solve the Dirac field equation in (1+1) space-time, and obtain the
relation between the current and the gauge field.  In Sec. IV, we
examine the gauge field degrees of freedom in two-dimensional (2D)
space-time. In Sec. V, we determine the equation of transverse
motion for fermions in a tube.  In Sec.\ VI, we present our
conclusions and discussions.

\section{$4D \to 2D$   Compactification in the Action Integral }

We employ the convention that before compactification is achieved,
all field quantities and gamma matrices are in four-dimensional
space-time unless specified otherwise.  With fermions interacting
with an SU(N) gauge field and a scalar field $m(x)$ in the (3+1)
Minkowski space-time, the SU(N) gauge invariant action integral
$\cal A$ is given by \cite{Pes95}
\begin{eqnarray}
&& {\cal A}   = \int d^4 x  \Bigg\{ Tr \Bigg[ {1\over 2} \left[
{\bar \Psi}  \gamma^\mu   \Pi_\mu \  \Psi
 - {\bar \Psi}  ~ m (x)~ \Psi  \right]
 - {1\over 2} \left[ {\bar \Psi}  \gamma^\mu
{\overleftarrow \Pi_\mu } \ \Psi   + {\bar \Psi}  ~ m (x) ~
 \Psi  \right] \Bigg] -
{1\over 4  } F^a_{\mu \nu}  \ F_a^{\mu \nu} \Bigg\}, \label{1}
\end{eqnarray}
where $A^a_\mu $ and $\Psi $ are the gauge and fermion fields
respectively in the Minkowski (3+1)-dimensional space-time with
coordinates $x\equiv x^\mu = (x^0 , {\bb x})=(x^0 , x^1, x^2,
x^3)$ and transverse coordinates $\bb r_\perp=(x^1, x^2)$. Here in
Eqs. (\ref{1}) ,
\begin{eqnarray}\label{eq2}
  \Pi_\mu &=&  i \partial_\mu +  g_{4D} ~ T_a A^a_\mu,
  \nonumber  \\
 F^a_{\mu \nu} &=&
\partial_\mu A^a_\nu  - \partial_\nu A^a_\mu  + g_{4D} ~ f^a_{~bc } ~ A^b_\mu ~ A^c_\nu  ,
 \\&\equiv& \partial_\mu
A^a_\nu - \partial_\nu A^a_\mu - i g_{4D} [ A^b_\mu ,~ A^c_\nu]^a
, \label{4}
\end{eqnarray}
$\gamma^{\nu}$ are the standard Dirac matrices, $\partial_{\mu} =
(\partial /\partial t , \nabla )$, and $a,b,c = 1 \dots N^2 -1$
are SU(N) group indices. We use the signature of (1, -1, -1, -1)
for the diagonal elements of metric tensor $ {g}_{\mu \nu}$.  We
should note that the action integral is gauge invariant since $
m(x)$ is independent of the SU(N) group generators $T_a$.

\subsection {Fermion part of the action integral}

The $4D$-action integral ${\cal A } $ resides in four-dimensional
(3+1) space-time.  There are environments in which the full
four-dimensional space-time is necessary, as for example in the
discussion of the phase transition in a hot quark-gluon plasma
\cite{Kar09,Fod04,Phi11}.  There are however environments which
are susceptible for compactification to two-dimensional (1+1)
space-time, in which the dynamics can be greatly simplified.

A proper environment for the compactification of QCD$_4$ can be
found in the special case in which a valence quark and antiquark
pull part from each other at high energies, as in the case
examined by Casher, Kugut, and Susskind \cite{Cas74}.  It is
convenient to work in the center-of-mass frame of the receding
quark-antiquark pair in which the magnitudes of the longitudinal
momenta of the valence quark pairs are very large, much larger
than the magnitudes of their transverse momenta.  Under such a
dominance of longitudinal dynamics, not only are the magnitudes of
the longitudinal momenta of the leading valence quark and
antiquark pair large, so are also those of the produced $q$ and
$\bar q$ partons.  It is then convenient to choose the Lorentz
gauge
\begin{eqnarray}
\partial ^\nu A_\nu^a=0.
\label{7A}
\end{eqnarray}
In this Lorentz gauge, $A_\nu^a$ is given by an integral of the
current $J_\nu^a$.  For a system with longitudinal dominance, the
magnitudes of the transverse currents are much smaller than the
magnitudes of the longitudinal currents.  As a consequence, the
magnitudes of the gauge field transverse components, $A_1^a$ and
$A_2^a$, along the transverse directions are small in comparison
with those of $A_0^a$ and $A_3^a$. The gauge field components
$A_1^a$ and $A_2^a$ can be neglected.  The absence of the
transverse components of the gauge fields in the Lorentz gauge
provides a needed simplification for compactification. However,
both $A_0^a$ and $A_3^a$ still depend on the $4D$ space-time
variables, $A_0^a ( x^0 , {\bb x} ), ~ A_3^a ( x^0 , {\bb x} )$.

The dominance of the longitudinal motion implies that the valence
leading quark and anti-quark lie inside a longitudinal tube.  The
limiting average transverse momentum suggests further that the
produced quarks reside within the longitudinal tube with a radius
inversely proportional to this limiting average transverse
momentum. As the confinement of the produced quarks within the
tube is a nonperturbative process that is beyond the realm of
perturbative QCD, we can represent the confinement property in
terms of a confining scalar interaction $S({\bb r}_\perp)$ in
transverse coordinates $\bb r_\perp$, with the quark mass function
$m(\bb r_\perp)=m_0+S({\bb
  r}_\perp)$.  The origin of ${\bb r}_\perp$ coordinates lies along the
longitudinal axis of the receding valence quark pair.  Because of
the presence of a scalar interaction $m(\bb r_\perp)$, our
dynamical problem does not maintain general Lorentz in all
directions. There remains however approximate Lorentz invariance
with respect to a finite boost along the longitudinal axis and the
range of this finite boost increases as the energy of the receding
quark pair increase.

Under such circumstances, we can carry out the compactification of
QCD$_4$ in (3+1) dimensions as follows.  The fermion part of the
$4D$-action ${\cal A } _F $ in (\ref{1}) is given by
\begin{eqnarray}\label{eq7}
 {\cal A }_F   \!=\!    Tr \!\! \int\! d^4 x \Biggl\{ \!{1\over 2}
{\bar \Psi}   \gamma^\mu\Pi_\mu   \Psi  \! -\! {1\over 2}  {\bar
\Psi}  \gamma^\mu{\overleftarrow \Pi_\mu }  \Psi  \!-\! {\bar
\Psi}   m ({\bb r}_\perp )
 \Psi  \!\Biggr\},~
\label{7}
\end{eqnarray}
where $\mu = 0,1,2,3$ and $ \gamma^\mu$ is
 the  $4D$-Dirac matrices,
\begin{eqnarray}\label{eq8}
&& \gamma^0   = \left( \begin{array}{cccc} 0 \
\ \ \ \ \  I  \\ \\
I \ \ \ \ \ \  0  \\
 \end{array} \right) , \ \ \ {\bb \gamma }  = \left( \begin{array}{cccc} 0 \
\ \ \ \ \  - {\bb \sigma}   \\ \\
{\bb \sigma} \ \ \ \ \ \  0
 \end{array} \right).
\end{eqnarray}
To relate the field variables in four-dimensional space-time to
those in 2-dimensional space-time in such a way that the
four-dimensional action integral can be simplified, we write the
Dirac fermion field $\Psi ( x )$ in terms of the following
bispinor with transverse functions $G_\pm({\bb r}_\perp)$ and
$x^0$-$x^3$ functions $f_\pm(x^0,x^3)$ \cite{Won91},
\begin{eqnarray}\label{eq9}
&& \Psi ( x ) \equiv \left( \begin{array}{cccc}  \varphi ( x^0 ;
{\bb x} )
  \\
\chi ( x^0 ; {\bb x} )
  \end{array} \right) \equiv
 \left( \begin{array}{cccc}  \varphi_1 ( x^0 ; {\bb x} )
  \\ \varphi_2 ( x^0 ; {\bb x} ) \\
\chi_1 ( x^0 ; {\bb x} ) \\ \chi_2 ( x^0 ; {\bb x} )
  \end{array} \right)
= \frac {1}{\sqrt{2}}\left( \begin{array}{cccc}  G_1  ({\bb
r}_\bot ) \left( f_+ (x^0 ; x^3 ) + f_- (x^0 ; x^3 ) \right) \
  \\
 - G_2  ({\bb r}_\perp) \left( f_+ (x^0 ; x^3 ) - f_- (x^0 ; x^3 )
\right)
 \\
 G_1  ({\bb r}_\perp) \left( f_+ (x^0 ; x^3 ) - f_- (x^0 ; x^3 )
\right)
 \\
  G_2  ({\bb r}_\perp) \left( f_+ (x^0 ; x^3 ) + f_- (x^0 ; x^3 )
\right)
  \end{array} \right) ,
\end{eqnarray}
where ${\bb r}_\perp$ is a vector in the plane  perpendicular to
the $x^3$ axis.  Using this explicit form of the Dirac bispinor
$\Psi$,  we can carry out  simplifications (with detailed
derivation given in Appendix A) that  lead from Eq.\ (\ref{7})
eventually to
\begin{eqnarray}
 {\cal A }_F   &=&   Tr~  \int d^2 X \Biggl\{ {1\over 2}
{\bar \Psi} (2D, X)\left[ i \gamma^\mu (2D) \partial_\mu + g_{2D}
\gamma^\mu T_a A^a_\mu (2D, X)\
\right] \Psi (2D, X)  \nonumber \\
&-& {1\over 2}  {\bar \Psi} (2D, X) \left[ i \gamma^\mu (2D)
\overleftarrow{\partial}_\mu - g_{2D} \gamma^\mu (2D) T_a A^a_\mu
(2D, X)\ \right]\Psi (2D, X)\nonumber\\
&-& {\bar \Psi} (2D, X) ~ m_{qT}  ~
 \Psi (2D, X) \Biggr\}\equiv  {\cal A }_F (2D)  ,~~~~~ \mu=0,3,
\label{10}
\end{eqnarray}
where we have introduced in the Dirac fermion field $\Psi(2D,X)$,
$\gamma$-matrices, and metric tensor $g_{\mu \nu}$, according to
the following specifications in the (1+1)-dimensional QCD$_2$
space-time
\begin{eqnarray}
&& \Psi ( 2D,X )= \left( \begin{array}{cccc} f_+ ( X ) \\
f_- ( X )
  \end{array} \right) , \ \ \ \ \ \ \ X = ( x^0 ; x^3 ) ,
  \label{11a}\\
  &&
\gamma^0 (2D)= \left( \begin{array}{cccc} 1\ \ \ \ \ \ 0 \\ \\
0
    \ \ \ \ \ -1 \\
 \end{array} \right) , \ \ \  \gamma^3 (2D)   = \left( \begin{array}{cccc} 0 \
\ \ \ \ \ 1 \\ \\ -1 \ \ \ \ \ \ 0
 \end{array} \right), ~~~~~ g_{\mu \nu }(2D)= \left( \begin{array}{cccc} 1 \ \ \ \ \ \ 0 \\ \\
0
    \ \ \ \ \ \ -1 \\
 \end{array} \right).
\label{11}
\end{eqnarray}
The $2D$ coupling constant, $g_{2D}$, is defined by the following
equation (see Appendix A)
\begin{eqnarray}
g_{2D}=\int dx^1 dx^2 g_{4D} [{ |G_1({\bb r}_\perp)|^2+|G_2({\bb
    r}_\perp)|^2}]^{3/2},
\label{13}
\end{eqnarray}
where the transverse wave functions $G_{1,2}({\bb r}_\perp)$ are
normalized according to
\begin{eqnarray}
\label{norm}   \int d x^1 d x^2 \left( \vert G_1 ({\bb r}_\perp)
\vert^2 +
  \vert G_2 ({\bb r}_\perp) \vert^2 \right) = 1 .
\label{14}
\end{eqnarray}
In the special case of the transverse ground state, we can
approximate the transverse density by a uniform distribution with
a sharp transverse radius $R_{T{\rm sharp}}$,
\begin{eqnarray}\label{eq15}
 \left( \vert G_1 ({\bb r}_\perp)
\vert^2 +
  \vert G_2 ({\bb r}_\perp) \vert^2 \right) \sim \frac{1}{\pi R_{T{\rm sharp}}^2}\Theta(R_{T{\rm sharp}}-|\bb r_T|),
\end{eqnarray}
we then obtain for a sharp distribution in the transverse ground
state the approximate relation  \cite{Won09}
\begin{eqnarray}\label{eq16}
g_{2D}\sim \frac{g_{4D}}{\sqrt{\pi} R_{T{\rm sharp}}}.
\end{eqnarray}

If we characterize the transverse ground state with a Gaussian
profile and a root-mean-square transverse radius
$R_T=\sqrt{2}\sigma_T$ as
\begin{eqnarray}
\label{Gauss} \left( \vert G_1 ({\vec r}_\bot ) \vert^2 +
 \vert G_2 ({\vec r}_\bot ) \vert^2 \right)&=&\frac{1}{2\pi \sigma^2} \exp\{ - \frac{r^2}{2\sigma_T^2}
 \},
\label{16a}
\end{eqnarray}
then  the  corresponding $g_{2D}$ coupling constant becomes
\begin{eqnarray}
 \label{g2D}
 g_{2D} &=& \frac{g_{4D}}{R_T}~ \sqrt{\frac{2}{9\pi}}.
\end{eqnarray}
The transverse quark mass  $m_{qT}$ in Eq.\ (\ref{10}) is given by
(see  Appendix A)
\begin{eqnarray}
 m_{qT} \!=\!\!  \int\!\! d x^1 d x^2  \left\{
m ({\bb r}_\perp)   \left( \vert G_1
  ({\bb r}_\perp) \vert^2 - \vert G_2 ({\bb r}_\perp) \vert^2
  \right) + \left( G^\ast_1 ({\bb r}_\perp) ( p_1 - ip_2 ) G_2 ({\bb  r}_\bot ) \right) - \left( G_1 ({\bb r}_\perp) ( p_1 + ip_2 )
  G^\ast_2 ({\bb r}_\perp) \right) \right\}\!.
\label{mqT}
\end{eqnarray}
The transverse quark mass $m_{qT}$ contains a contribution from
the quark rest mass (through $m(r)$), in addition to a
contribution arising from the confinement of the quark in the flux
tube (through the confining wave functions $G_{1,2}({\bb
r}_\perp)$). In obtaining these results, we have considered $2D$
gauge fields $A_\mu^a(2D,x^0,x^3)$ to be related to the $4D$-field
gauge fields $A_\mu^a(x^0,x^3,{\bb r}_\perp)$ by
\begin{eqnarray}
\label{AA} A_\mu^a(x^0,x^3,{\bb r}_\perp)&=& \sqrt{ |G_1({\bb
r}_\perp)|^2+|G_2({\bb
    r}_\perp)|^2} A_\mu^a(2D,x^0,x^3),~~~\mu=0,3.
\label{20}
\end{eqnarray}
The above equation means that along with the confinement of the
fermions, for which the wave function $G_{1,2}({\bb r}_\perp)$ is
confined within a finite region of transverse coordinates ${\bb
  r}_{\perp}$, the gauge field $A_\mu^a ( x)$, $\mu=0,3$, is also
considered to be confined within the same finite region of
transverse coordinates, as in the case for a flux tube.  Note that
because of the longitudinal dominance, we have assumed that $
A_\mu^a(x^0,x^3,{\bb
  r}_\perp)= 0$ for $\mu=1,2$.

\subsection{Gauge field part of the action integral}

Having reduced the fermion part of the action integral ${\cal
A}_F$, we come to examine the gauge field part of the action
integral ${\cal
  A}_A$,
\begin{eqnarray}
&& {\cal A}_A   = -\frac{1}{4}\int d^4 x
 F^a_{\mu \nu}  \ F_a^{\mu \nu} .
\label{19}
\end{eqnarray}
Our task is to find out  what will be the form of ${\cal A}_A$
involving  the gauge fields $A_\mu(2D)$ the two-dimensional
space-time, when $A_\mu(2D)$  the $A_\mu$ in four-dimensional
space-time are related  by Eq.\ (\ref{20}).

In Eq.\ (\ref{19}) the summation over $\mu,\nu$ includes terms
with $\mu,\nu=1,2$.  Previously, in going from ${\cal A}_F$ in
Eq.\ (\ref{7}) to ${\cal A}_F(2D)$ in the action integral of Eq.\
(\ref{10}), we have assumed that the currents in the $x^0$ and
$x^3$ directions are much greater in magnitude than the currents
in the transverse directions so that $A_1^a$ and $A_2^a$ are small
in comparison and can be neglected. As a consequence,
$F_{12}(4D)=0$ (we omit the superscript color index $a$ for
simplicity).

We consider now the contribution one of the terms, $F_{03}F^{03}$,
in Eq.\ (\ref{19}).  Equation (\ref{20}) gives
$F_{03}(x^0,x^3,{\bb
  r}_\perp)$ in four-dimensional space-time as
\begin{eqnarray}
F_{03}(x^0,x^3,{\bb r}_\perp) &=& [{ |G_1({\bb
r}_\perp)|^2+|G_2({\bb
    r}_\perp)|^2}]^{1/2}
[\partial_0 A_3(2D,x^0,x^3)-\partial_3 A_0(2D,x^0,x^3)]
\nonumber\\
& & - i g_{4D}  [{ |G_1({\bb r}_\perp)|^2+|G_2({\bb
    r}_\perp)|^2}] [A_0(2D,x^0,x^3),A_3(2D,x^0,x^3)]. \label{eq21}
\end{eqnarray}
On the other hand, the gauge field $F_{03}(2D,x^0,x^3)$ in two
dimensional space-time is given by definition as
\begin{eqnarray}\label{eq22}
F_{03}(2D,x^0,x^3)&=&\partial_0 A_3(2D)-\partial_3 A_0(2D)
-ig_{2D} [A_0(2D),A_3(2D)],
 \end{eqnarray}
where for brevity of notation, the coordinates $(x^0,x^3)$ in
$A_\mu (2D,x^0,x^3)$ will be understood.  As a consequence,
$F_{03}(2D,x^0,x^3)$ in two-dimensional space-time and
$F_{03}(x^0,x^3,{\bb r}_\perp)$ in four-dimensional space-time are
related by
\begin{eqnarray}\label{eq23}
F_{03}(x^0,x^3,{\bb r}_\perp)&=& [{ |G_1({\bb
r}_\perp)|^2+|G_2({\bb
    r}_\perp)|^2}]^{1/2}\{ F_{03}(2D,x^0,x^3)+ig_{2D} [A_0(2D),A_3(2D)]\}
\nonumber\\
& & - i g_{4D}  [{|G_1({\bb r}_\perp)|^2+|G_2({\bb
    r}_\perp)|^2}] [A_0(2D),A_3(2D)].
\end{eqnarray}
The above equation can be re-written as
\begin{eqnarray}\label{eq24}
F_{03}(x^0,x^3,{\bb r}_\perp) &=& [{ |G_1({\bb
r}_\perp)|^2+|G_2({\bb
    r}_\perp)|^2}]^{1/2}
\biggl \{ F_{03}(2D,x^0,x^3)
\nonumber\\
& &+\biggl [  i g_{2D} - i g_{4D}  [{ |G_1({\bb
r}_\perp)|^2+|G_2({\bb
    r}_\perp)|^2}]^{1/2}\biggr ] [A_0(2D),A_3(2D)] \biggr \}.
\end{eqnarray}
The product $F_{03} (x)F^{03}(x)$ in eq.\ (\ref{19}) becomes
\begin{eqnarray}\label{eq25}
& &F_{03}(x^0,x^3,{\bb r}_\perp) F^{03}(x^0,x^3,{\bb r}_\perp) =
[{ |G_1({\bb r}_\perp)|^2+|G_2({\bb
    r}_\perp)|^2}]\biggl \{ F_{03}(2D,x^0,x^3)F^{03}(2D,x^0,x^3)
\nonumber\\
&+& \biggl [  i g_{2D} - i g_{4D}  \{{( |G_1({\bb
r}_\perp)|^2+|G_2({\bb
    r}_\perp)|^2})\}^{1/2}\biggr ] \biggl (  F_{03}(2D)  [A^0(2D),A^3(2D)]
+ [A_0(2D),A_3(2D)]F^{03}(2D) \biggr )
\nonumber\\
& & +\biggl [ i g_{2D} - i g_{4D}  [{( |G_1({\bb
r}_\perp)|^2+|G_2({\bb
    r}_\perp)|^2})]^{1/2}\biggr ]^2\biggl (  [A_0(2D),A_3(2D)]
 [A^0(2D),A^3(2D)] \biggr )
\biggr \}. \label{26}
\end{eqnarray}
The action integral ${\cal A}_A$ in Eq.\ (\ref{19}) involves the
integration of the above quantity over $x_1$ and $x_2$. Upon
integration over $x^1$ and $x^2$, the second term inside the curly
bracket of the above equation, is zero,
\begin{eqnarray}
\label{second} & &\int dx^1 dx^2 [{( |G_1({\bb
r}_\perp)|^2+|G_2({\bb
    r}_\perp)|^2})]\biggl [ig_{2D} - ig_{4D} [{( |G_1({\bb r}_\perp)|^2+|G_2({\bb
    r}_\perp)|^2})]^{1/2}\biggr ]=0,
\label{25}
\end{eqnarray}
where we have used the relation between $g_{2D}$ and $g_{4D}$ as
given by Eq.(\ref{13}) and the normalization condition of
(\ref{norm}).  As a consequence, the integral of $F_{03}
(x)F^{03}(x)$ in Eq.\ (\ref{19}) becomes
\begin{eqnarray}
& &\int dx F_{03}(x^0,x^3,{\bb r}_\perp) F^{03}(x^0,x^3,{\bb
r}_\perp) =\int dx  [{ |G_1({\bb r}_\perp)|^2+|G_2({\bb
    r}_\perp)|^2}]\biggl \{ F_{03}(2D,x^0,x^3)F^{03}(2D,x^0,x^3)
\nonumber\\
& & +\biggl [ i g_{2D} - i g_{4D}  [{( |G_1({\bb
r}_\perp)|^2+|G_2({\bb
    r}_\perp)|^2})]^{1/2}\biggr ]^2\biggl (  [A_0(2D),A_3(2D)]
 [A^0(2D),A^3(2D)] \biggr )
\biggr \}.
\end{eqnarray}
For the second term in the curly bracket, the integral over $dx^1$
and $dx^2$ is
\begin{eqnarray}
\int dx^1 dx&^2  [{ |G_1({\bb r}_\perp)|^2+|G_2({\bb
    r}_\perp)|^2}]\biggl [ i g_{2D} - i g_{4D}  [{( |G_1({\bb
r}_\perp)|^2+|G_2({\bb
    r}_\perp)|^2})]^{1/2}\biggr ]^2
\end{eqnarray}
which can be considered as an integral over $g_{2D}$ in the form
\begin{eqnarray}
2 i\int dg_{2D}\int dx^1 dx&^2  [{ |G_1({\bb
r}_\perp)|^2+|G_2({\bb
    r}_\perp)|^2}]\biggl [ i g_{2D} - i g_{4D}  [{( |G_1({\bb
r}_\perp)|^2+|G_2({\bb
    r}_\perp)|^2})]^{1/2}\biggr ].
\end{eqnarray}
Because of Eq.\ (\ref{25}), the above integral gives an irrelevant
constant which we can set to zero.  After these manipulations, we
obtain
\begin{eqnarray}\label{eq29}
\int d x^1 d x^2 F_{03}(x^0,x^3,{\bb r}_\perp) F^{03}(x^0,x^3,{\bb
r}_\perp) &=& \int d x^1 d x^2
 [{( |G_1({\bb r}_\perp)|^2+|G_2({\bb
    r}_\perp)|^2})] F_{03}(2D,x^0,x^3)F^{03}(2D,x^0,x^3)
\nonumber\\
&=& F_{03}(2D,x^0,x^3)F^{03}(2D,x^0,x^3).
\end{eqnarray}
Following the same way (see Appendix B), we calculate terms
containing $F_{01}(4D)$, $F_{02} (4D)$, $F_{31} (4D)$, and $F_{32}
(4D)$.  For the gauge field part of the action integral, we obtain
\begin{eqnarray}
\frac{1}{4}  \int dx
 F^a_{\mu \nu} F_a^{\mu
\nu} &=&\frac{1}{4} \int {dx^0 dx^3}  F^a_{03 } (2D,x^0,x^3 )
 F_a^{03}(2Dx^0,x^3) \nonumber\\
&-& \int \frac{dx^0 dx^3}{ 4 } \int dx^1 dx^2 \biggl (
\{\partial_1[{ |G_1({\bb r}_\perp)|^2
   +|G_2({\bb
    r}_\perp)|^2}]^{1/2} \}^2
\nonumber\\
& &\hspace*{2.0cm}+\{\partial_2[{ |G_1({\bb r}_\perp)|^2+|G_2({\bb
    r}_\perp)|^2}]^{1/2} \}^2\biggr )
\nonumber\\
& &\times [A_0(2D,x^0,x^3)A^0(2D,x^0,x^3)
+A_3(2D,x^0,x^3)A^3(2D,x^0,x^3)]. \label{28a}
\end{eqnarray}
It is useful to introduce the gluon mass $m_{gT}$ that arises from
the confinement of the gluons in the transverse direction
\begin{eqnarray}\label{mgT}
 m_{gT}^2= \frac{1}{2} \int dx^1 dx^2
\biggl [ \{\partial_1[{ |G_1({\bb r}_\perp)|^2+|G_2({\bb
    r}_\perp)|^2}]^{1/2} \}^2+\{\partial_2[{ |G_1({\bb r}_\perp)|^2+|G_2({\bb
    r}_\perp)|^2}]^{1/2} \}^2\biggr ].
\end{eqnarray}
Equation (\ref{28a}) becomes
\begin{eqnarray}\label{eq32}
 \frac{1}{4} \int {d x}
 F^a_{\mu \nu} \ F_a^{\mu
\nu}&=&  \frac{1}{4} \int {dx^0 dx^3}
 \biggl \{ F^a_{03 } (2D) F_a^{03}(2D) - 2 m_{gT}^2  [A_0^a (2D) A^0_a(2D)
+A_3^a (2D)A^3_a(2D)] \biggr \}.
\end{eqnarray}
We collect all the fermion and gauge field parts of the action in
${\cal A} (4D )$ in Eq.\ (\ref{1}).  The action integral ${\cal
  A}={\cal A}_F+{\cal A}_A$ that was an integral in four-dimensional
space-time now turns into an integral  only in two-dimensional
space-time. All quantities in ${\cal A} (4D )$ are completely
defined in (1+1) dimensional space-time coordinates, we rename
this action integral ${\cal A} (2D )$ that is given explicitly by
\begin{eqnarray}
{\cal A} (2D ) = \int d^2 X &\Bigg\{&  Tr \Bigg[   {1\over 2}
\left[ {\bar \Psi}
    (2D,X) \gamma^k(2D) \Pi_k(2D)  \Psi (2D,X) - {\bar \Psi} (2D,X) m_{qT} \Psi (2D,X)
    \right]
\nonumber   \\
& &- {1\over 2} \left[ {\bar \Psi} (2D,X) \gamma^k(2D)
    {\overleftarrow \Pi_k }(2D) \ \Psi (2D,X) + {\bar \Psi} (2D,X) m_{qT} \Psi (2D,X)
    \right] \Bigg] \nonumber \\ && -{1\over 4 } F^a_{\mu \nu} (2D)
  \ F_a^{\mu \nu} (2D)  +{1\over 2 }m_{gT}^2 [A_a^\mu(2D)A^a_\mu(2D)]   \Bigg\},
\label{33}
\end{eqnarray}
where $\{\mu,\nu\}$=0,3, and
\begin{eqnarray}\label{eq34,35}
  \Pi_\mu (2D) &=&  i \partial_\mu +  g_{2D} ~ T_a A^a_\mu (2D, X)
= p_\mu  + g_{2D} ~ T_a A^a_\mu (2D, X) .
\end{eqnarray}
Here, all terms (including matrices and coefficients) in the
action integral of Eq.\ (\ref{33}) are in the $(1+1) $ Minkowski
space-time. Thus, in the environment of longitudinal dominance and
transverse confinement, we succeed in compactifying the action
integral in four-dimensional space-time to two-dimensional
space-time
 by judiciously relating  the field operators in four-dimensional space-time to the corresponding  field operators in two-dimensional space-time.

The result in this subsection indicates that the compactified
two-dimensional action integral has the same form as QCD in
two-dimensional space-time, and the compactified field theory  can
be appropriately call QCD$_2$. It has the feature that the coupling
constant $g_{2D}$  in QCD$_2$ acquires the dimension of a mass, and
is related to  $g_{4D}$ and the wave functions of the confined
fermions in the flux tube.  Fermions in different excited states
inside the tube will have different coupling constants as
indicated in Eq.\ (\ref{13}).  Furthermore, the action integral
gains additional transverse mass terms with an effective quark
mass $m_{qT}$ and gluon mass $m_{gT}$ that also depend on the
transverse fermion wave functions, as given in Eqs.\ (\ref{mqT})
and  (\ref{mgT})  respectively.  The transverse quark mass
includes a contribution from the quark rest mass, in addition to a
contribution due to the confinement of the flux tube. In the lower
two-dimensional space-time, fermions in excited transverse states
have a quark transverse mass different from those in the ground
transverse states.
 All the
transverse flux tube information is subsumed under these
quantities.

Provided that the fields $A_\mu^a(4D, x^1,x^2,x^0,x^3)$ are
governed by the standard gauge tarnsformation \cite{Pes95}, the 2D
gauged fields $A_\mu^a(2D,x^0,x^3)$ introduced according to Eq.\
(\ref{AA}) are found to transform under a gauge tarnsformation as
follows (see Appendix C) :
\begin{eqnarray}
\label{2DDtrans} &&A_\mu^a(2D,x^0,x^3) \to {\tilde
A}_\mu^a(2D,x^0,x^3) = A_\mu^a(2D,x^0,x^3) +
 f^{a}_{~b c} \varepsilon^b (x^0, x^3)
A_\mu^c (2D, x^0 ,x^3)
\end{eqnarray}
As a consequence, the term $A_\mu^a(2D)A^{\mu}_{ a} (2D)$
transforms under a gauge transformation as
\begin{eqnarray}
A_\mu^a(2D,x^0,x^3) A^\mu_a(2D,x^0,x^3) & \to & {\tilde
A}_\mu^a(2D,x^0,x^3) {\tilde A}^\mu_a(2D,x^0,x^3)  ,
\end{eqnarray}
 which indicates that the mass term in Eq.\ (\ref{33}) does not
 violate gauge invariance and does not violate the
 Slavnov-Taylor \cite{Slavnov} identities (see Appendix C)  due to the 2D gauge
 transformations given by Eqs.(\ref{2DDtrans}), (\ref{2Dtrans}).

\section{ Solution of the Dirac fields in (1+1) space-time}

Having completed the program of compactification of  QCD$_4$  to
QCD$_2$, we shall employ the new notation henceforth that all field
quantities and gamma matrices are in two-dimensional space-time
with $\mu=0,3$, unless specified otherwise.  We can use the QCD$_2$
action integral to get the equation of motion for the field.
Varying the action integral ${\cal A}(2D)$ given by Eq.(\ref{33})
with respect to $ {\bar \Psi}$, we derive the 2D Dirac equation,
\begin{eqnarray}
&& \left\{ i \gamma^{\mu} \left( \partial_{\mu} - i g_{2D} \cdot
A_{\mu}^a (X)~ T_a \right) - m_{qT} \right\} \Psi (X) = 0 ,
 \label{28}
\end{eqnarray}
where $2D$ Dirac matrices are those given in Eq.\ (\ref{11}). The
gauge field $A_{\mu}^a$ written in  component form is
\begin{eqnarray}\label{eq37}
&&  A_{\mu}^a  (X) = \left(  A^a_{0}  , -  A^a_{3} \right) ,
~~~~~~~ X = ( x^0 ; x^3 ) .
\end{eqnarray}
We express  the fermion field $\Psi $ in terms of $ f_+ (X)$ and
$f_- (X)$ as in Eq.\ (\ref{11a}),
\begin{eqnarray}\label{eq38}
&&    \Psi  =   \left(
\begin{array}{cccc}
1  \\
0
 \end{array} \right) ~f_+  (X) +    \left(
\begin{array}{cccc}
0  \\
1
 \end{array} \right) ~f_-  (X).
\end{eqnarray}
Then, the Dirac equation (37) becomes
\begin{eqnarray}\label{eq39}
 && i { \partial f_+ (t, z) \over \partial  t}  + i  {
\partial f_- (t, z) \over \partial  z} + g_{2D}\  T_a \ A_0^a f_+
 (t, z) - g_{2D}\  T_a \ A_3^a \ f_- (t, z)  =
m_{qT} f_+ (t, z),  \\
&&  - i { \partial f_- (t, z) \over \partial  t} - i  {
\partial f_+  (t, z) \over \partial  z} - g_{2D}\  T^a \ A^0_a \
f_- (z) + g_{2D}\  T_a \ A_3^a \ f_+ (t, z) = m_{qT} f_- (t, z) ,
\nonumber \\  && t\equiv x^0 ; ~~~ z\equiv x^3.
\nonumber \end{eqnarray}
We   introduce new functions as sum and difference of $f_+$ and
$f_-$:
\begin{eqnarray}\label{eq40}
&& \eta (t, z)= f_+ (t, z) - f_- (t, z) , \nonumber \\
&& \zeta (t, z)= f_+  (t, z) +  f_- (t, z).
 \end{eqnarray}
As a result, we obtain
\begin{eqnarray}
 && i { \partial \eta (t, z) \over \partial  t}  - i  {
\partial \eta (t, z) \over \partial  z} + g_{2D} {\hat A}_1 \eta (t, z)
= m_{qT}  \zeta(t, z), \nonumber \\
&& i { \partial \zeta (t, z) \over \partial  t}  + i  {
\partial \zeta (t, z) \over \partial  z} + g_{2D} {\hat A}_2 \zeta (t, z)  =
m_{qT}  \eta(t, z) , \label{42}
 \end{eqnarray}
where
\begin{eqnarray}\label{eq42}
&&   {\hat A}_1 = T_a ( A_0^a + A_3^a ); \ \ \ {\hat A}_2 = T_a (
A_0^a - A_3^a ).
  \end{eqnarray}
We look for a solution of Eq.\ (\ref{42})  in the form
\begin{eqnarray}\label{eq43}
&& \eta (t, z) = F (t, z )  \chi (t, z),
 \nonumber \\
&& \zeta (t, z) = G ( t, z ) \chi (t, z),
\end{eqnarray}
where the functions $\chi (t, z )$, $F (t, z )$ and $G (t, z )$
satisfy the following equations:
\begin{eqnarray}
 && i { \partial \chi (t, z) \over \partial  t}  - i  {
\partial \chi (t, z) \over \partial  z} + g_{2D} {\hat A}_1 \eta (t, z)
= 0, \nonumber \\
&& i { \partial \chi (t, z) \over \partial  t}  + i  {
\partial \chi (t, z) \over \partial  z} + g_{2D} {\hat A}_2 \zeta (t, z)  =
0  , \label{36}
 \end{eqnarray}
while
\begin{eqnarray}
 && i { \partial F (t, z) \over \partial  t}  - i  {
\partial F (t, z) \over \partial  z} = m_{qT}  G (t, z),
\nonumber \\
&& i { \partial G (t, z) \over \partial  t}  + i  {
\partial G (t, z) \over \partial  z}   =
m_{qT}  F (t, z). \label{37}
 \end{eqnarray}
The solution of Eq.\ (\ref{36}) can be formally written in the
operator form as follows:
\begin{eqnarray}\label{eq46}
&&\chi (t, z) =  \{T_{l (M_0 ; M) }\exp\} \left\{  i g_{2D} T_a
\int dx^\mu A^a_\mu \right\},
\end{eqnarray}
where the symbol $\{T_{l (M_0 ; M) }\exp\}$ means that the
integration is to be carried out along the line on the light cone
from the point $M_0$ to the point $M$ such that the factors in
exponent expansion are chronologically ordered from $M_0$ to $M$.
Eq.\ (\ref{37}) are the free 2D Dirac equations. When $m_{qT}$ is
a constant, the solution can be found as the superposition of 2D
plane waves:
\begin{eqnarray}\label{eq47}
 &&  F (t, z) = \int {d^2 P\over 2\pi} F (P) e^{\left( -i P X
 \right)}= \int {d^2 P\over 2\pi} F (P) e^{ -i (\omega t
-p_z z) },\nonumber \\
 &&  G (t, z) = \int {d^2 P\over 2\pi} G (P)e^{ \left( -i P X \right)}= \int {d^2 P\over 2\pi} G (P)e^{-i (\omega t
-p_z z) }.
  \end{eqnarray}
Substituting the last expansion into Eq.\ (\ref{37}),  we obtain
\begin{eqnarray}\label{eq48}
 &&  F (P) \left( \omega + p  \right) - m_{qT}  G (P) = 0 ,
 \nonumber \\
 &&  \left( \omega - p  \right) G (P)- m_{qT} \ F (P) = 0 ,
  \nonumber \\
 && P\equiv P^\mu = (\omega; {\bb p})= (\omega; p_z)\equiv(\omega; p).\label{48}
 \end{eqnarray}
As a result, we have
\begin{eqnarray}\label{eq49}
f_+ =\frac{\zeta+\eta}{2}\propto G(P)+F(p),
\nonumber\\
f_- =\frac{\zeta-\eta}{2}\propto G(P)-F(p).
\end{eqnarray}
Taking  $F(P)=1$ and  $G(P)=(\epsilon+p)/m_{qT}$, we derive
\begin{eqnarray}\label{eq50}
f_+ \propto \frac{\omega +p}{m_{qT}}+1,
\nonumber\\
f_- \propto \frac{\omega +p}{m_{qT}}-1.
\end{eqnarray}
The solution of Eq.\ (\ref{28})  becomes
\begin{eqnarray}\label{eq51}
  \Psi ( 2D, X ) &&= \Psi ( x^0 ; x^3 ) =   f_+  (
x^0 ; x^3 ) \left(
\begin{array}{c}
1  \\
0
 \end{array} \right) +   f_-  ( x^0 ; x^3 ) \left(
\begin{array}{c}
0  \\
1
 \end{array} \right)
\nonumber\\
&&= \int {d^2 P\over 2\pi} e^{ -i (\omega t -p_z z) } N(\omega,p)
\begin{pmatrix}
 \frac{\omega +p}{m_{qT}}+1 \cr
 \frac{\omega +p}{m_{qT}}-1\cr
\end{pmatrix} \cdot \chi (X) ,
\end{eqnarray}
where $N(\omega,p)$ is some normalization multiplier. We can take
the normalization condition
\begin{eqnarray}\label{eq52}
\left ( N(\omega,p)
\begin{pmatrix}
 \frac{\omega +p}{m_{qT}}+1 \cr
 \frac{\omega +p}{m_{qT}}-1\cr
\end{pmatrix}
 \right )^\dag
N(\omega,p)
\begin{pmatrix}
 \frac{\omega +p}{m_{qT}}+1 \cr
 \frac{\omega +p}{m_{qT}}-1\cr
\end{pmatrix}=\frac{1}{L},
\end{eqnarray}
where $L$ is the flux tube length.  Then, we obtain
\begin{eqnarray}\label{eq53}
N(\omega,p)
  =\frac{m_{qT}}{2\sqrt{L(\omega^2+p\omega})}.
\end{eqnarray}
As a result, the  general solution is
\begin{eqnarray}
 \Psi (X) &=&\int\limits_{-\infty}^{+\infty} \frac {d\omega}{2\sqrt L} \sum_p  {m_{qT}\ \over  \sqrt{\omega^2 + p \omega}} \
   \exp (-i P_\mu x^\mu ) \  a(p, \omega) \left[ \delta (\omega - \varepsilon (p)) + \delta (\omega + \varepsilon
   (p))\right] ~
\left(
\begin{array}{cccc}
{\omega + p \over m_{qT}} + 1  \\ \\ {\omega + p \over m_{qT}} - 1
 \end{array} \right)
 \nonumber\\
& \times & \{T_{l (M_0 ; M) }\exp\} \left\{  i g_{2D} T_a \int
dx^\mu A^a_\mu \right\}, \label{46a}
  \end{eqnarray}
where $a(p, \omega)$ are coefficients related to either particles
or anti-particles under the field quantization.  We have not
deliberately separated out positive and negative frequency terms
in Eq.\ (\ref{46a}) because the structure of the fermion vacuum is
strongly dependent on the explicit form of the external field
$A^a_\mu (X)$. Furthermore, when the external field depends on
time, there will be no stationary particles and antiparticles
states.

\subsection{Fermion current and gauge fields}

We envisage that a perturbative gauge field is introduced inside
the flux tube, such a field will generate a current, and the
current in turn will produce a gauge field self-consistently. How
do these quantities relate to each other? We therefore need to
obtain a relationship between the fermion current and the gauge
field.

The fermion field solution in Eq.\ (\ref{46a}) leads to a fermion
current $J_a^\mu$
\begin{eqnarray}
 &&  J_a^\mu (2D) = g_{2D}~Tr \left\{ {\bar \Psi ( X)} \gamma^\mu T_a  \Psi
 (X' )\right\}, \ \ \ \ \ X^\prime \to X.
\label{47a}
  \end{eqnarray}
Owing to the operation of  trace calculation in the last formula,
the current (\ref{47a}) contains the factor
\begin{eqnarray}
 &&
(T\exp) \left\{  i g_{2D}~ T_a   \int\limits_{X}^{X'}  A^a_\mu
dX^\mu \right\}.
\end{eqnarray}
We expand the operator exponent in the last equation as a series
with respect to  $( X' - X ) \to 0 $,
\begin{eqnarray}\label{eq57}
(T\exp) \left\{  i g_{2D}~ T_a   \int\limits_{X}^{X'}  A^a_\mu
dX^\mu \right\}&=& 1 + i g_{2D}~ T_a (X' - X )^\mu A^a_\mu (\xi) +
{i\over 2} g_{2D}~ T_a (X' - X )^\mu (X' - X )^\nu
\partial_\nu
A^a_\mu (\xi) \nonumber \\
& &~- g^2_{2D}~ ( T_a T_b ) ( {\tilde X}' - {\tilde X} )^\mu ( X'
- X )^\nu A^a_\mu ({\tilde \xi } ) A^b_\nu (\xi) \theta({\tilde
\xi } -  \xi  ), \label{49a}
\end{eqnarray}
where ${\tilde \xi } \in [ {\tilde X} , {\tilde X}' ],$ and $\xi
\in [  X ,
 X' ];~ \ X' \to X$.
We take the limits $( {\tilde X}' - {\tilde X})\to 0 $ and $( X' -
X )\to 0 $ such that
\begin{eqnarray}\label{eq58}
 && {( {\tilde X}' - {\tilde X} )\over  (  X' -
X )} \to 0.
\end{eqnarray}
Then, the last term in the expansion in Eq.\ (\ref{49a}) is equal
to zero. Substituting $(T\exp) \{ i g_{2D}~ T_a
\int\limits_{X}^{X'} A^a_\mu dx^\mu \}$ into Eq.(\ref{47a}), we
obtain for $(X'-X)\to 0$
\begin{eqnarray}\label{eq59}
J^\mu_a &&= \frac {g_{2D}}{ L} Tr \int d\omega \sum_{{ p}, f }
\Biggl\{ {<a^\dag_f ( p, \omega) a_f  (p, \omega)
> }( P^\mu T_a ) \left( - {\partial \over \partial P_\nu } \exp
\left( - i P (X' - X ) \right) \right) ~ \left[ \delta (\omega +
\varepsilon (p)) ~ + ~ \delta (\omega - \varepsilon (p))\right]
\nonumber \\
&&\times \left( g_{2D}~ T_b A^b_\nu (\xi) + {1\over 2} g_{2D}~ T_b
(X' - X )^\lambda
\partial_\nu
A^b_\lambda (\xi) \right)
 \Biggr\} ,
  \end{eqnarray}
where $f$ denotes flavor states. In reaching the last equation, we
have successively calculated a trace, gone from summation to
integration, introduced the additional integration with respect to
the $p$ variable, and integrated by parts. We note that upon
taking the partial derivative $\partial^2\equiv \partial_{(X)}^2$
on $(X'-X)^\lambda \partial_{\nu{(X)}} A^b_\lambda (\xi)$, we get
\begin{eqnarray}
\partial_{(X)}^2 \lim \limits_{X' \to~
X} \left\{ (X' - X )^\lambda \partial_{\nu(X)} A^b_\lambda (\xi )
\right\} &=&\lim \limits_{X' \to~ X}
\partial_{(X)}^2 \left\{ (X' - X )^\lambda
\partial_{\nu(X)}  A^b_\lambda(X)
 \right\}
\nonumber\\
&=&  \lim \limits_{X' \to~ X} \left\{ - 2
\partial_{(X)}^\lambda
\partial_{\nu(X)} A^b_\lambda  (X) + (X' - X)^\lambda
\partial_{\kappa(X)} ~ \partial_{(X)}^\kappa \{  \partial_{ \nu (X)} A^b_\lambda (X) \}\right\}.
  \end{eqnarray}
Upon taking the limit $X' \to X$, the second term vanishes .
Therefore, we have in the limit of $X' \to X$,
\begin{eqnarray}\label{cc1}
\lim \limits_{X' \to~ X} \left( (X' - X )^\lambda
\partial_\nu A^b_\lambda (\xi) \right) &=& - 2 \frac{\partial^\lambda
\partial_\nu}{
\partial^2} A^b_\lambda (X).
\end{eqnarray}

It should be noted that as QCD$_4$ in the (3+1) dimensional
space-time is gauge invariant, and we have chosen the Lorentz
gauge (\ref{7A}) in QCD$_4$  to simplify the compactified action
integral in QCD$_2$.  We need to continue to use the Lorentz gauge
in QCD$_2$ for consistency.  In the Lorentz gauge, the last term
in the second circular brackets in Eq.(\ref{eq59}) is equal to
zero because of Eq.(\ref{cc1}). Calculating a trace with respect
to the color variables according to Eq.(6), we can represent the
current $J_a^\mu$ in the following form
\begin{eqnarray}
 J_a^\mu(2D, X) &=& \frac { g^2_{2D}~ {\cal S}}{4}\     A_a^\mu
 (2D, X),
 \nonumber \\
 {\cal S} &=& \frac {1}{2\pi } \sum_f \int d^2 P \frac
{\partial}{\partial P^\mu} \left\{ \left[ \delta (\omega +
\varepsilon (p)) ~ + ~ \delta (\omega - \varepsilon (p))\right] ~
P^\mu ~ < a^\dag(p, \omega )_f ~a_f(p, \omega) > \right\}; ~~~ d^2
P = d\omega ~dp,
 \label{60}
  \end{eqnarray}
where $P^\mu$ is the  momentum introduced in Eq.\  (\ref{48}). We
can introduce a boson mass $m_{gf T}$ by
\begin{eqnarray}
m_{gf\,T}^2=  \frac{ g^2_{2D}{\cal S}}{4}.
\end{eqnarray}
Then, the current in Eq.(\ref{60}) can be written as
\begin{eqnarray}\label{SchCur }
 J_a^\mu(2D , X) &=& m_{gf\,T}^2  ~
   A_a^\mu(2D,X).
 \label{63}
  \end{eqnarray}
We can calculate the quantity ${\cal S}$. Changing $p$ by $- p$ in
the term corresponding to the negative $\omega$, we have
\begin{eqnarray}
 {\cal S} = \frac {1}{2\pi } \sum_f \int d^2 P \frac
{\partial}{\partial P^\mu} \left\{    P^\mu ~ \left( \delta
(\omega - \varepsilon (p)) < a^\dag(p, \omega)_f ~a_f(p, \omega) >
+ ~ \delta (\omega + \varepsilon (p)) < a^\dag(- p,  \omega)_f
~a_f( -p,  \omega) > \right) \right\}.
 \label{61}
  \end{eqnarray}
Integrating out the  $\delta$-functions, we obtain
\begin{eqnarray}
&& {\cal S} = \frac {1}{\pi } \sum_f   \int d p ~ \left( <
a^\dag_f ( p,  \varepsilon(p) ) ~a_f( p,  \varepsilon(p)) > + <
a^\dag_f (- p, - \varepsilon(p)) ~a_f( -p, - \varepsilon(p))
>\right) \left( 1  - \frac{m^2_{qT}}{2\varepsilon^2 (p)}\right).
 \label{62}
  \end{eqnarray}
Since  a fermion moves either along or opposite to  the only
spatial axis, we have
\begin{eqnarray}
&&  \int d p ~  < a^\dag_f ( p,  \varepsilon(p) ) ~a_f( p,
\varepsilon(p)) >  = \int d p ~ < a^\dag_f (- p, - \varepsilon(p))
~a_f( -p, - \varepsilon(p))
> = 1.
\label{62a}
  \end{eqnarray}
In the case of a flux tube for which $m_{qT} \gg p$, we obtain
\begin{eqnarray}
 {\cal S} &=&   \frac{2}{\pi}~N_f ,
  \end{eqnarray}
where $N_f$ is the number of flavors. We note in passing that in
the special case of the massless QED$_2$ \cite{Sch62}, we obtain
after summing over all spin states of a fermion
\begin{eqnarray}
&& {\cal S}_{\rm QED_2} = \frac {4}{\pi }, ~~(N_f =1),
 \label{62b}
  \end{eqnarray}
and
\begin{eqnarray}
&& m_{{gf~T}}^2  ({\rm QED_2})= \ \frac{g^2}{\pi},
  \end{eqnarray}
which agrees with the Schwinger massless QED$_2$ result
\cite{Sch62}.

Finally, we note that  under the gauge transformation
\begin{eqnarray}\label{eq64}
&& \delta A_a^\mu =  \varepsilon_b f_a^{\  bc} A_c^\mu
  \end{eqnarray}
the current (\ref{60}) satisfies the gauge relation
\begin{eqnarray}\label{eq65}
&& \delta J_a^\mu =  \varepsilon_b f^{~ b c }_a J_c^\mu.
  \end{eqnarray}

\section{Equation  of motion for the  2D Gauge Fields}

The action integral ${\cal A}$ allows us to obtain the equation of
motion for the 2D gauge field.  We rewrite the action integral
(\ref{33}) by expressing explicitly the term corresponding to the
interaction between the fermion and the gauge field,
\begin{eqnarray}
\label{QPA} {\cal A} (2D ) = \int d^2 X &\Bigg\{&  {i\over 2}
\left[ {\bar \Psi}
     \gamma^k \partial_k  \Psi  - {\bar \Psi}  m_{qT} \Psi
    \right]
- {i\over 2} \left[ {\bar \Psi}  \gamma^k
    {\overleftarrow \partial_k } \ \Psi  + {\bar \Psi}  m_{qT} \Psi
    \right] + J_a^\mu A_\mu^a \nonumber \\ && -{1\over 4 } F^a_{\mu \nu}
  \ F_a^{\mu \nu}    +{1\over 2 }m_{gT}^2  A^a_\nu A_a^\nu
   \Bigg\},
\end{eqnarray}
where $J_a^\mu(x)$ is the fermion current governed by
Eq.(\ref{63}).  Substituting $J_a^\mu(x)$ given by Eq.(\ref{63})
into the 2D action integral (\ref{QPA}), we obtain
\begin{eqnarray}
\label{QPA1} {\cal A} (2D ) = \int d^2 X &\Bigg\{&  {i\over 2}
\left[ {\bar \Psi}
     \gamma^k \partial_k  \Psi  - {\bar \Psi}  m_{qT} \Psi
    \right]
- {i\over 2} \left[ {\bar \Psi}  \gamma^k
    {\overleftarrow \partial_k } \ \Psi  + {\bar \Psi}  m_{qT} \Psi
    \right]  \nonumber \\ && -{1\over 4 } F^a_{\mu \nu}
  \ F_a^{\mu \nu}    +{1\over 2 }M_{gT}^2  A^a_\nu A_a^\nu
   \Bigg\}.
\end{eqnarray}
Here  the constant  $M_{gT}$  is given by
\begin{eqnarray} \label{M_gT}  M_{gT}^2& =& \frac{1}{2} \int
dx^1 dx^2 \biggl [ \{\partial_1[{ |G_1({\bb r}_\perp)|^2+|G_2({\bb
    r}_\perp)|^2}]^{1/2} \}^2+\{\partial_2[{ |G_1({\bb r}_\perp)|^2+|G_2({\bb
    r}_\perp)|^2}]^{1/2} \}^2\biggr ]  +  \frac{ g^2_{2D} {\cal S}}{2}     \nonumber \\
&  \equiv & m_{gT}^2 + m_{gf\,T}^2 \ge 0.
\end{eqnarray}

To find out the meaning of $M_{gT}$, we consider the variation of
the action integral (\ref{QPA1}) with respect to a variation of
the gauge field $ A_a^\nu (x)$.  We obtain equation of motion for
the variation $A_a^\nu (x)$. As a result, we derive the
Klein-Gordon-like equation:
\begin{eqnarray}
\label{KG}   \square ~  A_a^\nu &=& M_{gT}^2 ~ A_a^\nu.
\end{eqnarray}
 We look for a solution for the variation of the gauge field in
 Eq.\ (\ref{KG}) of the form
\begin{eqnarray}
\label{solKG}  && A_a^\nu = b_a (k, \nu )~ e_a^\nu (k)
\exp (- i k_\mu X^\mu ) , \nonumber \\ \nonumber \\
&&  k^\mu = (k^0 ; \bb k),~~ e_a^0 = \frac{\vert\bb k
\vert}{M_{gT}}~ (1, 0 ) , ~~~ e_a^3 = \frac{\vert k^0
\vert}{M_{gT}} (0, 1),
\end{eqnarray}
where $e_a^\nu$ denotes a pair orthogonal vectors; $b_a (k, \nu )$
are some coefficients being independent on $X$. Substituting $
A_a^\nu $ given by Eq.\ (\ref{solKG}) into Eq.\ (\ref{KG}), we
obtain:
\begin{eqnarray}\label{eq74}
 &&   (k^0 )^2 = \bb k^2  + M^2_{gT}.
  \end{eqnarray}
Because of both the positivity of $M^2_{gT}$ \  and Eq.
(\ref{eq74}),   $M_{gT}$ can be interpreted as a mass of the
particle whose energy is
\begin{eqnarray}\label{eq75}
 &&    E(k) \equiv k^0  = + \sqrt{  \bb k^2 +  M^2_{gT}} \label{spect}.
 \end{eqnarray}
Eqs.\ (\ref{solKG}) and (\ref{spect}) allow us to write down the
general solution of Eq.\ (\ref{KG}). Following the standard way
\cite{Pes95}, and separating the negative and positive frequency
terms, we obtain
\begin{eqnarray}
 A_a^\nu (2D, X)  &=&\sum_k  {e_a^\nu ~ M_{gT} \over  \sqrt{
\left( \bb k^2+ M^2_{gT}\right)^3 }}
  \Biggl\{
 \exp (-i k X) \  b_a(k,  \nu) ~ +  \exp ( + i k X) \  {\bar b}_a^\dag (k,  \nu)
\Biggr\} , \label{gensol-g}
\end{eqnarray}
where the symbols $b_a(k, \nu)$ and ${\bar b}_a^\dag (k, \nu)$ are
the operators of annihilation and creation of a boson with the
mass $M_{gT}$. In this way, $M_{gT}$ corresponds to the mass of
the boson responding to the space-time variation of the gauge
field variation.

The boson mass $M_{gT}$ in the action integral Eq.(\ref{QPA1})
arises from the compactification of $4D \to 2D$ and from the
interaction of the compactified fermions and gauge field. We
should also note that all masses (boson and fermion field) as well
as $2D$ coupling constant are governed by the functions of the
transverse motion of a fermion, $G_{1, 2} ({\vec r}_\bot )$. It
is independent of the color index $a$.  Out of the gauge field
variations of different color components $A_a^\mu$, one can
construct a colorless variation of the type
\begin{eqnarray}
A_{\rm color-singlet}^\nu = \frac{1}{\sqrt{8}}\sum_a A_a^\mu
|8,a\rangle,
\end{eqnarray}
where $|8,a\rangle$ is the color-octet state with component $a$.
Eq. (\ref{KG}) gives
\begin{eqnarray}
   \square ~  A_{\rm color-singlet}^\nu &=& M_{gT}^2 ~
A_{\rm color-singlet}^\nu .
\end{eqnarray}
Thus, we find that $M_{gT}$ is also the mass corresponding to a
colorless variation of the gauge field of different color
components in a flux tube.  Such a colorless variation should lead
to an observable quantity.  If one considers pion as the colorless
dynamical response of the variations of the gage fields in a
string, then $M_{gT}$ may be presumed to be the mass of the pion
within the environment of a flux tube under consideration.

\section{Equations of transverse motion in a tube
and the Fermion effective mass}

To obtain the equations of motion for the functions $G_1 ({\vec
  r}_\bot )$ and $G_2 ({\vec r}_\bot )$, we vary the action integral
${\cal A} (4D)$ in Eq.\ (\ref{1}) with the fermion fields $\Psi
(4D , x)$ given by Eq.(\ref{eq9}), under the constraint of the
normalization condition Eq.\  (\ref{norm}). To do this we
construct a new functional ${\cal F}$,

\begin{eqnarray}\label{eq80}
&& {\cal F} = {\cal A }(4 D ) + \frac{ \lambda}{2} ~ \int d x^1 d
x^2 \left( \vert G_1 ({\vec r}_\bot ) \vert^2  + \vert G_2 ({\vec
r}_\bot ) \vert^2 \right) ~\int d x^0 d x^3 \left( {\bar \Psi}
(x^0 , x^3 )
 ~ { \Psi}  (x^0 , x^3 ) \right),
\end{eqnarray}
where $\lambda$ is the Lagrange multiplier. The last term in Eq.\
(\ref{eq80}) takes into account the unitarity of a fermion field
in the 4D  space-time.  Varying the last equation with respect to
the functions $G_1 ({\vec r}_\bot )$ and $G_2 ({\vec
  r}_\bot )$, we obtain
\begin{eqnarray}\label{eq81}
&&
( p_1 +  ip_2 )  G_1 ({\vec r}_\bot )  = (  m ({\vec r}_\bot ) +\lambda )  G_2  ({\vec r}_\bot ), \nonumber \\
&& ( p_1 - ip_2 ) G_2 ({\vec r}_\bot ) = (\lambda   - m  ({\vec
r}_\bot ) )
G_1  ({\vec r}) ,\nonumber \\
&&( p_1 +  ip_2 )  G^\ast_2 ({\vec r}_\bot )  = (  m ({\vec r}_\bot ) - \lambda )  G^\ast_1  ({\vec r}_\bot ), \nonumber \\
&& ( p_1 - ip_2 ) G^\ast_1 ({\vec r}_\bot ) = - (  m ({\vec
r}_\bot ) + \lambda ) G^\ast_2 ({\vec r}).
\end{eqnarray}
Carrying out complex conjugation in the   last   two equations, we
obtain
\begin{eqnarray}\label{eq82}
&& \lambda = \lambda^\ast.
\end{eqnarray}
Combining Eq.\ (\ref{eq81}), we get
\begin{eqnarray}\label{eq83}
&& \left( p^2_1 +  p^2_2  - \lambda^2 + m^2  ({\vec r}_\bot )
\right)  G_1 ({\vec r}_\bot )  = G_2 ({\vec r}_\bot ) ( p_1 - ip_2
)  m  ({\vec r}_\bot )
 \nonumber \\
&& \left( p^2_1 +  p^2_2  - \lambda^2 + m^2  ({\vec r}_\bot )
\right)  G_2 ({\vec r}_\bot )  = - G_1 ({\vec r}_\bot ) ( p_1 +
ip_2 )  m  ({\vec r}_\bot ).
\end{eqnarray}
Substituting the equations (\ref{eq81}) for $G_{1,2} ({\vec
r}_\bot )$ functions into the formula (\ref{mqT}) for $m_{qT}$, we
find that
\begin{eqnarray}\label{eq84}
 m_{qT}& = & \lambda .
\end{eqnarray}
Thus, the effective mass of the compactified 2D fermion field is
equal to the energy eigenvalue for the transverse motion of the 4D
fermion as described in Eqs. (\ref{eq81}). We should note here
that the 2D fermion can generally gain a mass even when the
initial 4D fermion appears to be massless.  The compactification
effectively leads to a constraint in moving a fermion from one
point of a space-time to another point due to decreasing the
number of trajectories in the 2D space-time as compared with the
4D situation.  This contraint leads to the presence of an
effective mass.

\section {Conclusions and Discussions}

Encouraged by the successes of the particle production model of
Casher, Kogut, and Susskind using the Abelian gauge field theory
in two-dimension space-time \cite{Cas74} and the Lund model of
string fragmentation \cite{And83}, we seek a compactification of
QCD$_4$ to QCD$_2$ in the environment of a flux tube.  Under the
assumption of longitudinal dominance and transverse confinement,
the SU(N) gauge invariant field theory of QCD$_4$ can be
compactified in the (1+1) Minkowski space-time, from the
consideration of the action integral. This is achieved by finding
a way to relate the field variables in 2-dimensional space-time to
those in four-dimensional space-time.

The compactified 2D action integral ${\cal A} (2D)$ depends only
on fields that are defined in two-dimensional space-time.  It has
the same structure  as those in QCD in four-dimensional space-time
and can therefore be appropriately called QCD$_2$.  In the
compactified QCD$_2$ quantum field theory, the coupling constant
is found to be dimensional, and there are additional terms in the
action associated with an effective quark mass and effective gauge
field mass as a result of the flux tube confinement.  These
quantities depends on the transverse profile and the transverse
state of the quarks in the flux tube.

On a basis of the derived QCD$_2$ action integral, the equations
of motion for the fields can be obtained for both the fermion
field and the gauge field.  The solution of 2D Dirac equation can
then be formally obtained. The structure of the solution allows
one to consider the effects of the fermion-gluon coupling.  As a
result, the 2D action integral can be re-written in the form such
that the gauge field acquires an additional effective mass due to
interaction with fermions.  The structure of the derived mass term
appears to be identical to the one obtained by Schwinger
\cite{Sch62} in the special case of massless QED$_2$.

The occurrence of a massive composite bound state in gauge field
theories has been known in many previous investigations
\cite{xxx}. How the massive bosons as a pole in the three gluon
vertex in 4-dimensional space-time \cite{Pap11,Pap12} can be
produced in the flux tube environment in high-energy collisions
will be an interesting subject worthy of further investigations.

\vspace*{0.3cm}

\centerline{\bf Acknowledgment}

\vspace*{0.3cm}   The research  was supported in part by the
Division of Nuclear Physics, U.S. Department of Energy.

\appendix
\section{}
Substituting $\Psi (4D, x)\equiv  \Psi ( x)$ given by Eq.\ (7)
into the first term in Eq.\ (5), we obtain
\begin{eqnarray}\label{a1}
&& {\bar  \Psi (x)} \gamma^\mu   \Pi_\mu \  \Psi (x) = \chi^\dag
\left( \Pi_0 - ( p_1 \sigma_1 + p_2 \sigma_2 ) - \Pi_3 \sigma_3
\right) \chi + \varphi^\dag \left( \Pi_0 + ( p_1 \sigma_1 + p_2
\sigma_2 ) + \Pi_3 \sigma_3 \right) \varphi \nonumber \\ \nonumber \\
&& = \chi^\dag \left[ \left( \begin{array}{cccc} \Pi_0 - \Pi_3 \
\ \ \ \ \  0  \\ \\
0 \ \ \ \ \ \  \Pi_0 + \Pi_3  \\
 \end{array} \right) - \left( \begin{array}{cccc} 0 \
\ \ \ \ \  p_1 - i p_2   \\ \\
p_1 + i p_2  \ \ \ \ \ \  0
 \end{array} \right)
\right] \chi + \varphi^\dag \left[ \left( \begin{array}{cccc}
\Pi_0 + \Pi_3 \
\ \ \ \ \  0  \\ \\
0 \ \ \ \ \ \  \Pi_0 - \Pi_3  \\
 \end{array} \right) + \left( \begin{array}{cccc} 0 \
\ \ \ \ \  p_1 - i p_2   \\ \\
p_1 + i p_2  \ \ \ \ \ \  0
 \end{array} \right)
\right] \varphi   \nonumber \\
&&= \chi_1^\ast (\Pi_0 - \Pi_3 ) \chi_1 + \chi_2^\ast (\Pi_0 +
\Pi_3 ) \chi_2 - \chi_1^\ast ( p_1 - ip_2 ) \chi_2 - \chi_2^\ast (
p_1 + ip_2 ) \chi_1 + \varphi_1^\ast (\Pi_0 + \Pi_3 ) \varphi_1 +
\varphi_2^\ast (\Pi_0 - \Pi_3 ) \varphi_2   \nonumber \\
&&~~~~~+ \varphi_1^\ast ( p_1 - ip_2 ) \varphi_2 + \varphi_2^\ast
( p_1
+ ip_2 ) \varphi_1  \nonumber \\
&&= \chi_1^\ast (\Pi_0 - \Pi_3 ) \chi_1 + \chi_2^\ast (\Pi_0 +
\Pi_3 ) \chi_2  + \varphi_1^\ast (\Pi_0 + \Pi_3 ) \varphi_1 +
\varphi_2^\ast (\Pi_0 - \Pi_3 ) \varphi_2   \nonumber \\
&&~~~~~ - \chi_1^\ast ( p_1 - ip_2 ) \chi_2 - \chi_2^\ast ( p_1 +
ip_2 ) \chi_1 + \varphi_1^\ast ( p_1 - ip_2 ) \varphi_2 +
\varphi_2^\ast ( p_1 + ip_2 ) \varphi_1.
\end{eqnarray}
Integration of the last equation gives
\begin{eqnarray}\label{a2}
\int d^4 x \ {\bar  \Psi (x)} \gamma^k   \Pi_k \  \Psi (x) &=&
\int d^4 x \left\{  \chi_1^\ast (\Pi_0 - \Pi_3 ) \chi_1 +
\chi_2^\ast (\Pi_0 + \Pi_3 ) \chi_2  + \varphi_1^\ast (\Pi_0 +
\Pi_3 ) \varphi_1 + \varphi_2^\ast (\Pi_0 - \Pi_3 ) \varphi_2
\right\}  \nonumber\\
& & +
 \int d^4 x \left\{    -  \chi_1^\ast ( p_1 - ip_2 ) \chi_2 -
\chi_2^\ast ( p_1 + ip_2 ) \chi_1 + \varphi_1^\ast ( p_1 - ip_2 )
\varphi_2 + \varphi_2^\ast ( p_1 + ip_2 ) \varphi_1  \right\}
\nonumber \\
&=& \int d^4 x \left\{  \chi_1^\ast (\Pi_0 - \Pi_3 ) \chi_1 +
\chi_2^\ast (\Pi_0 + \Pi_3 ) \chi_2  + \varphi_1^\ast (\Pi_0 +
\Pi_3 ) \varphi_1 + \varphi_2^\ast (\Pi_0 - \Pi_3 ) \varphi_2
\right\} \nonumber \\
&& -   \int d^4 x \left\{   G^\ast_1 ({\vec r}_\bot ) ( p_1 - ip_2
) G_2 ({\vec r}_\bot ) \right\} \left( \vert f_+ \vert^2 - \vert
f_- \vert^2 \right)  \nonumber \\
&=&  \int d^4 x \left( \vert G_1  ({\vec r}_\bot )  \vert^2  +
\vert G_2 ({\vec r}_\bot ) \vert^2  \right) \left[ \ f^\ast_+
\Pi_0 f_+ + \ f^\ast_- \Pi_0 f_- +  \ f^\ast_+ \Pi_3 f_-  + \
f^\ast_- \Pi_3 f_+ \right] \nonumber \\
&& -  \int d^4 x \left\{ G^\ast_1 ({\vec r}_\bot ) ( p_1 - ip_2 )
G_2 ({\vec r}_\bot ) \right\} \left( \vert f_+ \vert^2 - \vert f_-
\vert^2 \right).
\end{eqnarray}
Following the same way, we derive for the term ${\bar \Psi} (x)
\gamma^k {\overleftarrow \Pi_k } \ \Psi (x) $
\begin{eqnarray}\label{a3}
 \int d^4 x \ {\bar \Psi (x)} \gamma^k {\overleftarrow\Pi}_k \ \Psi
 (x) &=&  \int d^4 x \left( \vert G_1 ({\vec r}_\bot ) \vert^2 +
 \vert G_2 ({\vec r}_\bot ) \vert^2 \right) \left[ \ f^\ast_+
   {\overleftarrow\Pi}_0 f_+ + \ f^\ast_- {\overleftarrow\Pi}_0 f_- +
   \ f^\ast_+ {\overleftarrow\Pi}_3 f_ - +f^\ast_-
      {\overleftarrow\Pi}_3 f_+ \right] \nonumber \\ & & -  \int d^4
 x \left\{ G_1 ({\vec r}_\bot ) ( p_1 + ip_2 ) G^\ast_2 ({\vec r}_\bot
 ) \right\} \left( \vert f_+ \vert^2 - \vert f_- \vert^2 \right).
\end{eqnarray}
We substitute $ \Psi (x)$ of Eq.\ (7) into the last term in Eq.\
(5), and  we obtain
\begin{eqnarray}\label{a4}
&& {\bar \Psi} (x) m ({\vec r}_\bot ) \Psi (x) =   m ({\vec
r}_\bot
  ) \ \left( \vert G_1 ({\vec r}_\bot ) \vert^2 - \vert G_2 ({\vec
    r}_\bot ) \vert^2 \right) \left[ \vert f_+ \vert^2 - \vert f_-
    \vert^2 \right].
\end{eqnarray}
Collecting the above results and introducing the $2D$-fermion wave
function $\Psi (X)$, $2D$-gamma matrices $\gamma^\mu$, and the
metric tensor $g_{\mu \nu}(2D)$ as given in Eqs.\ (\ref{11a}) and
(\ref{11}), we obtain the Fermion part of the action integral in
Eq.\ (8).

\section{}
To compactify the gauge field parts of the (3+1) dimensional
space-time to (1+1) dimensional space-time, we need to evaluate
$F_{01}$, $F_{02}$,
 $F_{31}$, and $F_{32}$. Direct calculations give (color indexes
 are omitted for simplicity)
\begin{eqnarray}\label{b1}
F_{01}(x^0,x^3,{\bb r}_\perp)&=& -\partial_1 A_0( x^0,x^3,{\bb
r}_\perp)
\nonumber\\
&=&-\partial_1[{ |G_1({\bb r}_\perp)|^2+|G_2({\bb
    r}_\perp)|^2}]^{1/2} A_0(2D,x^0,x^3),
\end{eqnarray}
\begin{eqnarray}\label{b2}
F_{01}( x^0,x^3,{\bb r}_\perp) F^{01}( x^0,x^3,{\bb r}_\perp)
&=&\{-\partial_1[{ |G_1({\bb r}_\perp)|^2+|G_2({\bb
    r}_\perp)|^2}]^{1/2} \}\{-\partial^1[{ |G_1({\bb r}_\perp)|^2+|G_2({\bb
    r}_\perp)|^2}]^{1/2} \}
\nonumber\\
& & \times A_0(2D,x^0,x^3)A^0(2D,x^0,x^3),~~~
\nonumber\\
&=&- \{\partial_1 [{ |G_1({\bb r}_\perp)|^2+|G_2({\bb
    r}_\perp)|^2}]^{1/2} \}^2
A_0(2D,x^0,x^3)A^0(2D,x^0,x^3),
\end{eqnarray}
which contribute a gauge field mass in the $A_0(2D,x^0,x^3)$ gauge
field.   Similarly,  we can calculate
\begin{eqnarray}\label{b3}
&&F_{02}( x^0,x^3,{\bb r}_\perp) F^{02}( x^0,x^3,{\bb r}_\perp) =
-\{\partial_2[{ |G_1({\bb r}_\perp)|^2+|G_2({\bb
    r}_\perp)|^2}]^{1/2} \}^2 A_0(2D,x^0,x^3)A^0(2D,x^0,x^3),
    \nonumber \\ \nonumber \\
&& F_{31}( x^0,x^3,{\bb r}_\perp) F^{31}( x^0,x^3,{\bb r}_\perp) =
-\{\partial_1[{ |G_1({\bb r}_\perp)|^2+|G_2({\bb
    r}_\perp)|^2}]^{1/2} \}^2 A_3(2D,x^0,x^3)A^3(2D,x^0,x^3),
\end{eqnarray}
which contribute a gauge field mass in the $A_3(2D,x^0,x^3)$ gauge
field.  Similarly, we have also
\begin{eqnarray}
F_{32}( x^0,x^3,{\bb r}_\perp) F^{32}( x^0,x^3,{\bb r}_\perp)
&=&-\{\partial_2[{ |G_1({\bb r}_\perp)|^2+|G_2({\bb
    r}_\perp)|^2}]^{1/2} \}^2 A_3(2D,x^0,x^3)A^3(2D,x^0,x^3).
\end{eqnarray}
Combining all similar terms, we get
\begin{eqnarray}\label{b4}
& &[F_{01}F^{01}+F_{02}F^{02}+F_{31}F^{31} +F_{32}F^{32}](
x^0,x^3,{\bb r}_\perp)
\nonumber\\
&=&-\biggl ( \{\partial_1[{ |G_1({\bb r}_\perp)|^2+|G_2({\bb
    r}_\perp)|^2}]^{1/2} \}^2+\{\partial_2[{ |G_1({\bb r}_\perp)|^2+|G_2({\bb
    r}_\perp)|^2}]^{1/2} \}^2\biggr )
\nonumber\\
& &\times [A_0(2D,x^0,x^3)A^0(2D,x^0,x^3)
+A_3(2D,x^0,x^3)A^3(2D,x^0,x^3)]
\nonumber\\
&=& -\biggl ( \{\partial_1[{ |G_1({\bb r}_\perp)|^2+|G_2({\bb
    r}_\perp)|^2}]^{1/2} \}^2+\{\partial_2[{ |G_1({\bb r}_\perp)|^2+|G_2({\bb
    r}_\perp)|^2}]^{1/2} \}^2\biggr )
\nonumber\\
& &\times [A_0(2D,x^0,x^3)A^0(2D,x^0,x^3)
+A_3(2D,x^0,x^3)A^3(2D,x^0,x^3)].
\end{eqnarray}
Then, due to the normalization relation (\ref{norm}) we have the
following (see Eq.\ (26) and (27)) for the gauge field part:
\begin{eqnarray}\label{b5}
& &{1\over 4 } \int d^4 x  F^a_{\mu \nu}(4D) \ F_a^{\mu \nu}(4D)
={1\over 4 } \int dx^0 dx^3 \int dx^1 dx^2 ( |G_1({\bb
r}_\perp)|^2+|G_2({\bb
  r}_\perp)|^2)  F^a_{03 } (2D)
 F_a^{03}(2D) \nonumber\\
& &  -{1\over 4 } \int dx^0 dx^3 \int dx^1 dx^2 \biggl (
\{\partial_1[{ |G_1({\bb r}_\perp)|^2+|G_2({\bb
    r}_\perp)|^2}]^{1/2} \}^2+\{\partial_2[{ |G_1({\bb r}_\perp)|^2+|G_2({\bb
    r}_\perp)|^2}]^{1/2} \}^2\biggr )
\nonumber\\
& &~~~~~~~~~~~\times [A_0(2D,x^0,x^3)A^0(2D,x^0,x^3)
+A_3(2D,x^0,x^3)A^3(2D,x^0,x^3)]
\nonumber\\
 &=& {1\over 4 } \int dx^0 dx^3  F^a_{0
  3} (2D) \ F_a^{0 3}(2D) - {1\over 2 } \int dx^0 dx^3
m_{gT}^2[A_0(2D)A^0(2D) +A_3(2D)A^3(2D)],
\end{eqnarray}
where $m_{gT}^2$ is the mass term that arises from the confinement
of the gluons in the transverse direction
\begin{eqnarray}\label{b6}
 m_{gT}^2= \frac{1}{2} \int dx^1 dx^2
\biggl [ \{\partial_1[{ |G_1({\bb r}_\perp)|^2+|G_2({\bb
    r}_\perp)|^2}]^{1/2} \}^2+\{\partial_2[{ |G_1({\bb r}_\perp)|^2+|G_2({\bb
    r}_\perp)|^2}]^{1/2} \}^2\biggr ].
\end{eqnarray}
Note that using integration by parts, we get
\begin{eqnarray}\label{b7}
& & -\int dx^1 dx^2
 \{\partial_1[{ |G_1({\bb r}_\perp)|^2+|G_2({\bb
    r}_\perp)|^2}]^{1/2} \}^2
\nonumber\\
&=& \int dx^1 dx^2 [{ |G_1({\bb r}_\perp)|^2+|G_2({\bb
    r}_\perp)|^2}]^{1/2} \partial_1^2
[{ |G_1({\bb r}_\perp)|^2+|G_2({\bb
    r}_\perp)|^2}]^{1/2}.
\end{eqnarray}
Therefore, adding the terms together, we obtain:
\begin{eqnarray}\label{b8}
& & m_{gT}^2= \frac{1}{2} \int dx^1 dx^2 \biggl [ \{\partial_1[{
|G_1({\bb r}_\perp)|^2+|G_2({\bb
    r}_\perp)|^2}]^{1/2} \}^2
+\{\partial_2[{ |G_1({\bb r}_\perp)|^2+|G_2({\bb
    r}_\perp)|^2}]^{1/2} \}^2 \biggr ]
\nonumber\\
&=& - \frac{1}{2} \int dx^1 dx^2 [{ |G_1({\bb
r}_\perp)|^2+|G_2({\bb
    r}_\perp)|^2}]^{1/2}( \partial_1^2 +  \partial_2^2)
[{2 |G_1({\bb r}_\perp)|^2+|G_2({\bb
    r}_\perp)|^2}]^{1/2}
\nonumber\\
&=& \int dx^1 dx^2 [{ |G_1({\bb r}_\perp)|^2+|G_2({\bb
    r}_\perp)|^2}]^{1/2} \left( - \frac{1}{2} \nabla_T^2 \right)
[{ |G_1({\bb r}_\perp)|^2+|G_2({\bb
    r}_\perp)|^2}]^{1/2} .
\end{eqnarray}

\section{}

\subsection{Transformation of a gauge field in the 2D space-time}

We would like to write down the gauge transformation properties
for $A_\mu^a (2D, x^0 , x^3)$.  For the corresponding gauge field
$A_\mu^a (x)$ in the 4D space-time $x=(x^0, x^3,{\bb r}_\perp)$,
it transforms under a gauge transformation as
 \cite{Pes95}
\begin{eqnarray}
\label{2Dtrans} && A_\mu^a(x)  \to {\tilde A}_\mu^a(x) =
A_\mu^a(x) +\delta A_\mu^a ( x).
\end{eqnarray}
where
\begin{eqnarray}
&& \delta A_\mu^a ( x) =  f^{a}_{~b c} \varepsilon^b (x) A_\mu^c (
x)  - \frac{1}{g(4D)} \partial_\mu \varepsilon^a (x)
\label{cgauge}
\end{eqnarray}
According to Eq.\ (\ref{AA}) the gauge fields in the 2D and 4D
space-time are related to each other as follows:
\begin{eqnarray}
\label{2} && A_\mu^a(2D,x^0,x^3) = \frac{ A_\mu^a(x^0,x^3,{\bb
r}_\perp)}{\sqrt {|G_1({\bb r}_\perp)|^2+|G_2({\bb
    r}_\perp)|^2}},~~~\mu=0,3,\nonumber \\ \nonumber \\
    && A_\mu^a(x^0,x^3,{\bb r}_\perp) =0  , ~~ \mu =1,2
\end{eqnarray}
From the last equation we have
\begin{eqnarray}
\label{3} &&\delta  A_\mu^a(x^0,x^3,{\bb r}_\perp) =0  , ~~ \mu
=1,2. ~~~~  \Rightarrow ~~ \partial_\mu \varepsilon^a (x) = 0 , ~~
\mu =1,2.
\end{eqnarray}
Then, we have
\begin{eqnarray}
&& \varepsilon^a (x) = \varepsilon^a (x^0 ,x^3)
\end{eqnarray}
Next, we would like to transform $A_\mu^a(2D,x^0,x^3)$ by using
the first relation  in Eq.(\ref{2}).
\begin{eqnarray}
\label{5} &&\delta A_\mu^a(2D,x^0,x^3) =  \frac{\delta
A_\mu^a(x^0,x^3,{\bb r}_\perp)}{\sqrt {|G_1({\bb
r}_\perp)|^2+|G_2({\bb
    r}_\perp)|^2}} + A_\mu^a(x^0,x^3,{\bb r}_\perp) \delta \left(\frac{1
}{\sqrt {|G_1({\bb r}_\perp)|^2+|G_2({\bb
    r}_\perp)|^2}}\right)
\end{eqnarray}
The last term in Eq.(\ref{5}) is equal to zero since
$\varepsilon^a = \varepsilon^a (x^0 ,x^3)$.  Then, substituting
Eq.\ (\ref{cgauge}) into Eq.\ (\ref{5}) we obtain
\begin{eqnarray}
\label{c6} &&\delta A_\mu^a(2D,x^0,x^3) =
 f^{a}_{~b c} \varepsilon^b (x^0, x^3)
A_\mu^c (2D, x^0 ,x^3)  - \frac{1}{g(4D)~\sqrt {|G_1({\bb
r}_\perp)|^2+|G_2({\bb
    r}_\perp)|^2}} \partial_\mu
\varepsilon^a (x^0,x^3)
\end{eqnarray}
Since the left-hand side of Eq.\ (\ref{c6})  depends
 on $(x^0,x^3)$ the same must be for the right-hand side of this equation.
This means that $\varepsilon^a (x^0,x^3)$= constant  and the
transformation relation for $ A_\mu^a(2D,x^0,x^3)$ is
\begin{eqnarray}
&&\delta A_\mu^a(2D,x^0,x^3) =
 f^{a}_{~b c} \varepsilon^b (x^0, x^3)
A_\mu^c (2D, x^0 ,x^3)
\end{eqnarray}
Varying  the mass term in the 2D Lagrangian with respect to the
group variables, we obtain
\begin{eqnarray}
\delta {\cal L}_{m_{gT}} &=& {1\over 2 }m_{gT}^2 \delta
[A_a^\mu(2D)A^a_\mu(2D)] = m_{gT}^2 \delta
[A_a^\mu(2D)]~[A^a_\mu(2D)]
\nonumber\\
&=&  m_{gT}^2
 f^{a}_{~b c} \varepsilon^b (x^0, x^3)
[A_\mu^c (2D)] [A_a^\mu(2D)]= 0,
\end{eqnarray}
due to the anti-symmetry of the structure constants $ f^{a}_{~b
c}$. Using Eq.\ (\ref{cgauge}) for the infinitesimal
transformation of the gauge of the field $A_\mu^a(2D,x^0,x^3)$ we
calculate the $n$-th variation of  $A_\mu^a(2D,x^0,x^3)$. After
such calculations we derive that the gauge transformation of the
2D gauge field has the form
\begin{eqnarray}
\label{8} \delta^{(n)} { A}_\mu^a(2D,x^0,x^3) &=& f^{a}_{~b c}
\varepsilon^b (x^0, x^3) f^{c}_{~~b_1 c_1} \varepsilon^{b_1} (x^0,
x^3)  A_\mu^{c_1} (2D, x^0 ,x^3) \dots f^{c_{n-2}}_{~~~b_{n-1}
c_{n-1}} \varepsilon^{b_{n-1}} (x^0, x^3)  A_\mu^{c_{n-1}} (2D,
x^0 ,x^3), \nonumber \\ {\tilde A}_\mu^a(2D,x^0,x^3) &=&
 e^{f^{a}_{~b c} \varepsilon^b (x^0, x^3)}~
A_\mu^c (2D, x^0 ,x^3), \nonumber \\
{\tilde A}^\mu_a(2D,x^0,x^3) &=& A_\mu^c (2D, x^0 ,x^3)~  e^{-
f^{a}_{~b c} \varepsilon^b (x^0, x^3)}.
\end{eqnarray}
As a consequence,
\begin{eqnarray}
{\tilde A}_\mu^a (2D, x^0 ,x^3) ~ {\tilde A}^\mu_a(2D,x^0,x^3) = {
A}_\mu^a (2D, x^0 ,x^3)~ {A}^\mu_a(2D,x^0,x^3),
\end{eqnarray}
which maintains the 2D gauge invariance of the derived 2D action
integral Eq.\ (\ref{33}).

\subsection{The Slavnov-Taylor identities in the 2D space-time in the Lorentz gauge}

The Slavnov-Taylor identities in the standard 4D space-time in the
Lorentz gauge has the form \cite{Slavnov}:
\begin{eqnarray}
\label{10AB} \int \exp \left( i {\cal A} (A)  \right) \cdot  \int
dz \left( J^\nu_b (z)
\partial_\nu (M^{-1})^{ba} (z,y) + g_{4D} f_{d~c}^{~b} ~ J_\nu^d (z) ~
A_b^\nu (z) ~ (M^{-1})^{ca} (z,y) \right) \Delta (A) d A = 0 ,
\end{eqnarray}
where $J_\nu^a (z)$ is a fermion current, $(M^{-1})^{ca} (z,y)$ is
the propagator of a scalar field, and $\Delta (A)$ is the
Faddeev-Popov determinate. After the compactification with respect
to Eqs.\ (\ref{eq9}), (\ref{norm}), and (\ref{AA}), the action
integral ${\cal A} (A)$ becomes ${\cal A}[2D, A(2D)]$. Integrating
the first term in the circular brackets by parts with respect to
the $z$ variable and using Eqs.(\ref{eq9}) and (\ref{AA}), we
obtain
\begin{eqnarray}
\label{10c}  \int dz \left( J^\nu_b (z)
\partial_\nu (M^{-1})^{ba} (z,y) \right)& =& - \int dz   (M^{-1})^{ba} (z,y)
\nonumber \\
& &  \times \Biggl( [{ |G_1({\bb z}_\perp)|^2+|G_2({\bb
    z}_\perp)|^2}]^{2}(\partial_0 J^0_b (z^0 ,z^3)
+ \partial_3 J^3_b (z^0 ,z^3))
\nonumber \\
& & ~~~-
     i Tr \left\{(p_1 - ip_2 ) G_1^\ast ({\bb
z}_\perp) G_2 ({\bb z}_\perp) {\bar \Psi }(z^0 ,z^3) T_b \Psi (z^0
,z^3) \right\} \Biggr).
\end{eqnarray}
The integral involving the first term inside the above curly
bracket is equal to zero because of the Lorentz gauge for the
$A^\mu_a$ field and Eq.(\ref{SchCur })
 while the second one is found to be the same due to the trace calculation.
Thus the first term in the circular brackets in Eq.\ (\ref{10AB})
is equal to zero.

As for the second term in the circular brackets in Eq.\
(\ref{10AB}), by using  Eqs.\ (\ref{norm}), (\ref{AA}), and
(\ref{SchCur }) it can be written as
\begin{eqnarray}
\int dz & &\left(  g_{4D} f_{d~c}^{~b} ~ J_\nu^d (z) ~
A_b^\nu (z) ~ (M^{-1})^{ca} (z,y) \right)   \nonumber \\
&=& g_{4D} f_{d~c}^{~b} \int dz [{ |G_1({\bb
z}_\perp)|^2+|G_2({\bb
    z}_\perp)|^2}]^{3/2}
  J_\nu^d (2D, z^0 ,z^3) ~ A_b^\nu (2D, z^0 ,z^3) ~ (M^{-1})^{ca}
(z,y)  \nonumber \\
& = &  m_{gf\,T}^2  ~ g_{4D} f_{d~c}^{~b} \int dz [{ |G_1({\bb
z}_\perp)|^2+|G_2({\bb
    z}_\perp)|^2}]^{3/2} A^d_\nu (2D, z^0 ,z^3)
   ~ A_b^\nu (2D, z^0 ,z^3) ~ (M^{-1})^{ca}(z,y) = 0.
\end{eqnarray}
The last expression is equal to zero because of the anti-symmetry
of the structure constant. Thus, the pre-exponent in Eq.\
(\ref{10AB}) is found to be equal to zero after the $4D \to 2D $
compactification. This means that the Slavnov-Taylor identities
are not violated in the 2D space-time we have considered.

\end{document}